\documentclass[preprint2]{aastex}

\usepackage{amsmath}

\usepackage{lineno}

\begin{document}


\title{The Effect of Lower Mantle Metallization on Magnetic Field Generation in Rocky Exoplanets}


\author{R. Vilim\altaffilmark{1}, S. Stanley\altaffilmark{1}}
\affil{Department of Physics, University of Toronto, Toronto, Ontario, Canada}
\and
\author{L. Elkins-Tanton\altaffilmark{2}}
\affil{Department of Terrestrial Magnetism, Carnegie Institution for Science, Washington, D.C., USA}
\email{rvilim@physics.utoronto.ca}


\begin{abstract}
Recent theoretical and experimental evidence indicates that many of the materials that are thought to exist in the mantles of terrestrial exoplanets will metallize and become good conductors of electricity at mantle pressures. This allows for strong electromagnetic coupling of the core and the mantle in these planets. We use a numerical dynamo model to study the effect of a metallized lower mantle on the dynamos of terrestrial exoplanets using several inner core sizes and mantle conductivities.

We find that the addition of an electrically conducting mantle results in stronger core--mantle boundary fields because of the increase in magnetic field stretching. We also find that a metallized mantle destabilizes the dynamo resulting in less dipolar, less axisymmetric poloidal magnetic fields at the core--mantle boundary. The conducting mantle efficiently screens these fields to produce weaker surface fields. We conclude that a conducting mantle will make the detection of extrasolar terrestrial magnetic fields more difficult, while making the magnetic fields in the dynamo region stronger.
\end{abstract}





\section{Introduction}
There is evidence on every planet in our solar system, except Venus, of a dynamo-generated planetary magnetic field either today, or at some point in the past. The existence and morphology of these dynamo-generated fields are strongly constrained by the properties of the planetary deep interior. With the vast array of extrasolar rocky planets now being discovered, an obvious question is what to expect for their dynamo-generated magnetic fields. Since these exoplanets provide a larger range of planetary properties than those found in our solar system, the potential exists for interior dynamics not seen in our solar system.

Detectability is an important factor in the study of extrasolar planetary magnetic fields. There are two ways that the magnetic field of an extrasolar planet could be detected from Earth. The first occurs when electrons from the stellar wind interact with the dynamo-generated magnetic field from the planet, emitting cyclotron radiation \citep{farrell1999, greissmeier2007, lecacheux1991}. The radiated power associated with this is 
\begin{equation}
P_{\mathrm{rad}}\propto B^{0.58} a^{-1.17}
\end{equation}
where $B$ is the magnetic field and $a$ is the planet--star distance. The constant of proportionality is related to the strength of the solar wind \citep{farrell1999}.

The second method by which extrasolar planetary magnetic fields could be detected is through magnetospheric interactions between a close-in planet and its host star. This mechanism was proposed to explain the observations of ``hot spots'' in the chromospheres of some stars, which rotate with the planetary orbit. If a planet with a magnetic field orbits close enough to a star, it is possible that the magnetic fields lines may join the two bodies and trap plasma in the closed field lines between them \citep{cohen2009}. The presence of the planetary magnetic field can be detected indirectly through interaction of this plasma with the host star.

Both of these signatures of extrasolar planetary magnetic fields are more easily observable when the magnetic field at the planetary surface is stronger.

Many of the large number of extrasolar terrestrial planets which have been discovered have a mass greater than that of Earth \citep{schneider2012}. This implies that the pressures and temperatures inside ``super-Earths'' can be considerably higher than in any of the terrestrial planets in our solar system \citep{valencia2006}. This raises the possibility of novel material properties inside the deep interiors of these planets that could potentially have an important effect on a planet's geodynamics.


One interesting possibility is the pressure-induced metallization of mantle materials. In the Earth, the mantle is largely electrically insulating, mainly because its main constituent (perovskite) is not expected to metallize until pressures far beyond any which are expected to be found in the deep interiors of even the largest rocky exoplanets \citep{tsuchiya2011}. In exoplanets, the possibility of compositions which differ significantly from Earth could lead to electrically conducting lower mantles in even small rocky exoplanets \citep{ohta2012}. Recent studies have shown that there are several common minerals which should metallize. These include CaSiO$_{3}$ \citep{tsuchiya2011}, FeO \citep{ohta2012}, and Al$_{2}$O$_{3}$ \citep{nellis2010} indicating that lower mantle metallization may be a phenomenon that is common in rocky exoplanets.

When mantle materials are metallized a number of their physical properties change significantly. A metallized mantle should have a high thermal conductivity as well as a high electrical conductivity because of the Wiedemann-Franz law, which states that the thermal conductivity is proportional to the electrical conductivity and temperature. Another concern is that the slope of the liquidus of the material could change, potentially leading to a liquid lower mantle. This can be definitively ruled out in the case of FeO, where experimental data show no change in the curvature of the liquidus when FeO metallizes \citep{boehler1992}. There are no experimental data on the melting temperatures of CaSiO$_{3}$ and Al$_{2}$O$_{3}$;  however, Schreinmaker's rule \citep{zen1966} implies that a liquidus with $dT/dP<0$ is not possible. Any melting requires that metallic CaSiO$_{3}$ and Al$_{2}$O$_{3}$ behave differently than FeO, and that a fortuitous combination of adiabat slope and liquidus slope exists. We feel this situation unlikely so we henceforth assume that the mantle is completely solid in this study.


\cite{chan2008} published numerical dynamo simulations with a conducting mantle layer that had  conductivity which varied sinusoidally in latitude and longitude. They found that the addition of this electrically conducting mantle layer could cause a previously steady dynamo to vacillate and, if the conducting mantle layer was thick enough, could stop a dynamo from operating altogether. However, the parameters this study chose were for the benchmark dynamo, an intentionally placid dynamo solution run at unrealistic parameters that is normally used to ensure that a numerical dynamo model is working correctly. This raises concerns about its relevance to planetary regimes since the parameters in the benchmark dynamo are much less realistic than those used in most planetary studies.

The purposes of these studies were not to model the metallization of the mantle, and so they used lower mantle conductivities than one might expect from a metallic mantle. Furthermore, they concentrated on the effects of heterogeneity in the mantle conductivity. Here we use a numerical dynamo model running at more realistic parameters to study the effect of lower mantle metallization on the dynamos of possible extrasolar terrestrial planets with specific interest in the observable properties of these dynamos.

The study of terrestrial exoplanets, especially terrestrial exoplanetary interiors remains underconstrained. For a dynamo to exist on these planets, among the most important factors is the state of the mantle. The power  to drive the dynamo of an extrasolar terrestrial planet is controlled by the mantle, so the efficient transfer of heat out of the planet is of great importance. The presence of plate, tectonics and vigorous mantle convection are both efficient ways that planets can drive dynamos in their cores. Currently, mantle convection on these bodies is poorly understood, especially for large exoplanets. The field remains sharply divided \citep{lenardic2012, oneill2007, stein2011, stein2013, valencia2009, vanheck2011} as to the likelihood of plate tectonics, and the viscosity structure \citep{karato2011} of large terrestrial exoplanets. 

The effect of a metallized mantle layer on mantle convection has been studied by \cite{vandenberg2010}. They found that the addition of an electrically conducting layer at the the bottom of the mantle caused the bottom boundary layer to heat, become buoyant and rise to the upper mantle. This leads to an increased heat flux at the core--mantle boundary (CMB) which could increase the power available to the dynamo, but shorten its lifetime. This would be a secondary effect as the heat flux increase is less than an order of magnitude. There are other, less well constrained properties of extrasolar terrestrial planetary mantles which will have a greater impact on the heat flux from the core than mantle metallization (e.g. radiogenic heating, or the presence of plate tectonics). 

\section{Expected Effects of Mantle Metallization on the Dynamo} 

An electrically conducting mantle should affect the dynamo in two ways. First, any quickly varying components of the magnetic field should be screened out by the skin effect before they reach the surface, weakening any observed field. Inside the solid mantle layer, the magnetic field obeys a diffusion equation, with a diffusivity equal to $\eta=1/(\sigma_{M} \mu_{o})$ where $\sigma_{M}$ is the conductivity of the layer and $\mu_{o}$ is the magnetic permeability of free space. Neglecting the spherical geometry of the core, the magnetic field is attenuated in a solid conducting mantle proportional to $e^{-d\sqrt{\omega/(2 \eta)}}$, where $\omega$ is the frequency of the magnetic field variations at the top of the core and $d$ is the thickness of the conducting mantle. This implies that the thicker the mantle layer, the weaker the observed field. Also, higher multipoles will be preferentially damped due to their more rapid time variation compared to lower multipoles \citep{christensen2004}, while the dipole component of the magnetic field may not be greatly affected.

This screening effect also applies to non-axisymmetric components of the field that are being advected by the background flow at the top of the core. For example, in dynamo simulations, equatorial flux spots are a common occurrence, and typically drift westward \citep{finlay2003}.  From the mantle reference frame these spots are viewed as a time varying magnetic field and hence, they will be screened. In cases where the dynamo generated field is exceptionally steady and axisymmetric, the screening effect may be unimportant as the timescale of field change will be very large.

The second feature we expect when a conducting mantle is added to a dynamo is magnetic shear at the core--mantle boundary due to flux freezing in the mantle. The magnetic field should anchor itself in both the solid mantle and the convecting liquid outer core \citep{moffatt1978}. Shear should then be created between the solid mantle and the strong zonal flows which are present in planetary cores. Simultaneously, the fluid in the outer core should feel a Lorentz force from the stretching of magnetic field lines anchored in the mantle. While the screening effect is a kinematic process that simply acts on time varying magnetic fields generated by other means, the magnetic coupling effect requires a fully coupled modeling approach.

The thickness of the metallized part of the mantle should have a significant effect on the character of the observable field. We first note that the screening effect and the Lorenz feedback effect scale differently with metallized mantle thickness. Electromagnetic screening becomes more important as the thickness of the metallic mantle layer increases, since the field is attenuated by distance proportional to $e^{-d\sqrt{\omega/(2 \eta)}}$. Conversely, the Lorenz force acting on the fluid by the mantle is nearly independent of mantle thickness. This is evident when examining the equation for the torque ($\Gamma$) on the core by the mantle (adapting \cite{gubbins1987})
\begin{equation}
\Gamma=\frac{r^{2}}{\mu_{o}}\iint\limits_{\mathrm{CMB}}B_r\left(\mathbf{r}\mathbf{\mathbf{\times}}\mathbf{B}\right)r^{2} \sin\theta d\theta d\phi
\end{equation}
The surface integral is over the CMB and does not involve mantle thickness. 

The thickness of the metallized mantle layer in a given exoplanet depends strongly on the size of the planet, the size of the core, and the material which is metallized. For example, FeO should metallize at approximately 55 GPa along the Earth's geotherm \citep{ohta2012} meaning that a planet significantly smaller than Earth could have an electrically conducting mantle layer as long as it had a large amount of FeO in the mantle. Conversely, CaSiO$_{3}$ is not expected to metallize until pressures of 600 GPa, meaning that a conducting mantle layer should form in CaSiO$_{3}$ rich planets with masses greater than 5$M_{\earth}$ \citep{tsuchiya2011}. By varying the composition, planet size, and core size, a conducting mantle layer of nearly any thickness can be achieved. Because of this we avoid fixing a planetary radius or mass, as the arguments here apply equally well to a range of planet sizes. As the surface magnetic field strength is strongly dependent on planetary size and core-mass fraction, we discuss only the properties of the field at the top of the electrically conducting region.
 
\section{Numerical Model}
To study the effect of an electrically conducting mantle on dynamo generation in planets we use the Kuang-Bloxham numerical dynamo model \citep{kuang1999}. This model has been successfully applied to study the magnetic field of many of the planets in our solar system. It numerically solves the three-dimensional, nonlinear, Boussinesq, magnetohydrodynamic equations in a convecting, rotating, spherical shell, representing the fluid outer core. The inner core of our model is solid, and conducts electricity with the same conductivity as the outer core.

In the model, all the equations are solved in their non-dimensional forms. We use the radius of the core ($r_{o}$) as the length scale, the magnetic diffusion time $\tau=r_{o}^{2}/\eta$ as the time scale and the magnetostrophic balance intensity $B=\sqrt{2\Omega\rho/\sigma_{C}}$ (where $\Omega$ is the rotation rate of the planet, $\rho$ is the density of the core, and $\sigma_{C}$ is the electrical conductivity of the outer core) as the magnetic field scale. We also use $h_{B} r_{o}$ as the temperature scale, where $h_{B}$ is the buoyancy flux at the inner core boundary. The non-dimensional equations of the model are:
\begin{eqnarray}
E_{\eta}\left(\frac{\partial}{\partial t}+\mathbf{u}\cdot\mathbf{\nabla}\right)\mathbf{u}+\mathbf{\hat{z}}\mathbf{\times}\mathbf{u}&=&-\mathbf{\nabla} p+\mathbf{J}\mathbf{\times}\mathbf{B}+Ra\Theta\mathbf{r}+E\mathbf{\nabla}^{2}\mathbf{u} \label{eq:momentum}\\
\frac{\partial T}{\partial t}+\mathbf{u}\cdot\mathbf{\nabla} T&=&q_{\kappa}\mathbf{\nabla}^{2}T \label{eq:temperature}\\
\frac{\partial \mathbf{B}}{\partial t}&=&\mathbf{\nabla}\mathbf{\times}\left(\mathbf{u}\mathbf{\times}\mathbf{B}\right)\label{eq:mie}+\mathbf{\nabla}^{2}\mathbf{B}
\label{eq:magnetic}
\end{eqnarray}
Here $\mathbf{B}$, $\mathbf{u}$, and $\mathbf{J}$ are the magnetic field, velocity and current density respectively, $p$ is the modified pressure, and $\hat{z}$ is the rotation axis of the system. Also, $T$ is the temperature, while $\Theta$ is the temperature perturbation.

The nondimensional numbers in Equations (\ref{eq:momentum})--(\ref{eq:magnetic}) are the Rayleigh number ($Ra$), the Ekman number ($E$), the magnetic Ekman number ($E_{\eta}$), and the Roberts number ($q_{\kappa}$), and are given by 
\begin{eqnarray}
Ra&=&\frac{\alpha_{T}g_{o}h_{B}r_{o}^{2}}{2\Omega\eta}\\
E&=&\frac{\nu}{2\Omega r_{o}^2}\\
E_{\eta}&=&\frac{\eta}{2\Omega r_{o}^{2}}\\
q_{k}&=&\frac{\kappa}{\eta}
\end{eqnarray}
where $\alpha_{T}$ is the thermal expansion coefficient, $g_{o}$ is the gravitational acceleration at the CMB, $\nu$ is the kinematic viscosity, and $\kappa$ is the thermal diffusivity. In all our models we set $Ra=12000$, $E=2.1125\times 10^{-5}$, $E_{\eta}=4.225\times 10^{-6}$ and $q_{k}=5$.

Our model differs from most numerical dynamo models in its treatment of the outer boundary. We include a solid mantle layer, the electrical conductivity of which can be specified arbitrarily. We specify the relative conductivity of the mantle layer with $\sigma_{MC}=\sigma_{M}/\sigma_{C}$ where $\sigma_{M}$ is the conductivity of the conducting mantle layer.

Our models use spherical harmonics as basis functions in the azimuthal directions, and a combination of Chebyshev expansions and compact finite differences in the radial direction. All models presented here have $L_{\mathrm{max}}=56$, $m_{\mathrm{max}}=53$ and the number of grid points in the inner core, outer core, and mantle are $N_{i}=50$, $N_{o}=64$, and $N_{m}=50$ respectively. For numerical reasons we make use of scale-dependent viscosities and diffusivities, we have applied them lightly in this study, using them only for $L>20$ in order to minimize their dynamical effects.

\begin{figure}
\plotone{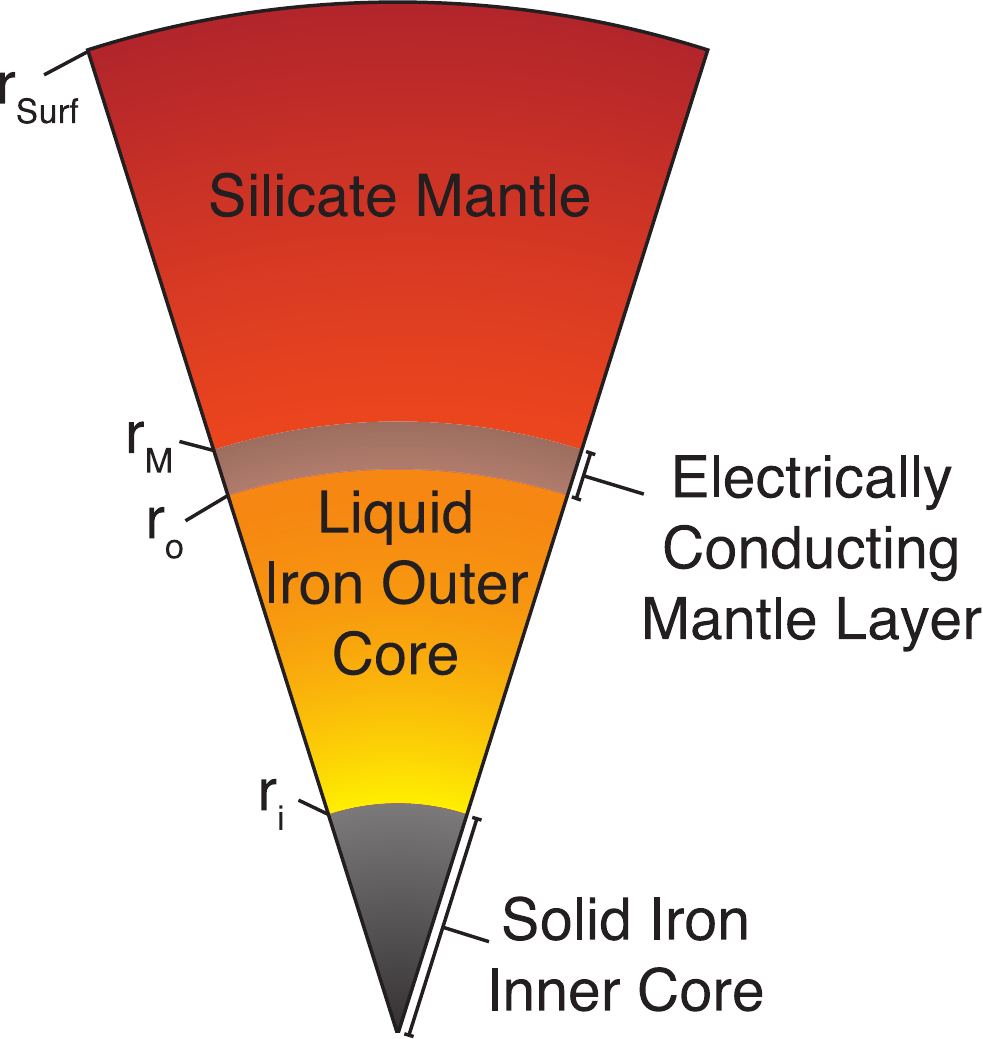}
\caption{Schematic diagram of the structure of a planet with a conducting mantle layer. Our numerical model solves in the region below $r_{M}$.}
\label{fig:structure}
\end{figure}

We model a metallized mantle with a spherical shell of uniform conductivity on the outside of the dynamo region. In all models the spherical shell extends from $r_{o}$ to 1.07$r_{o}$. We also make the shell highly conducting, varying $\sigma_{MC}$ from $1/8$ to $1$. As a control, we run a model with a relatively insulating mantle ($\sigma_{MC}=1/400$). A schematic diagram of the model is shown in Figure \ref{fig:structure}.

The strong rotational influence on convection in planetary cores implies that the solid inner core has a significant effect on the dynamo \citep{heimpel2005, stanley2007}. Since exoplanets can be found in  different stages of their thermal evolution (depending on their age), we use three different inner core sizes in this study. As a planet ages, the heat lost from the core will cause the solid inner core to grow, so the different cases can be considered proxies for different stages of the life of the dynamo. Here we use $r_{io}=0.35$, $0.5$, and $0.70$.

\section{Results}
\begin{figure}
\plotone{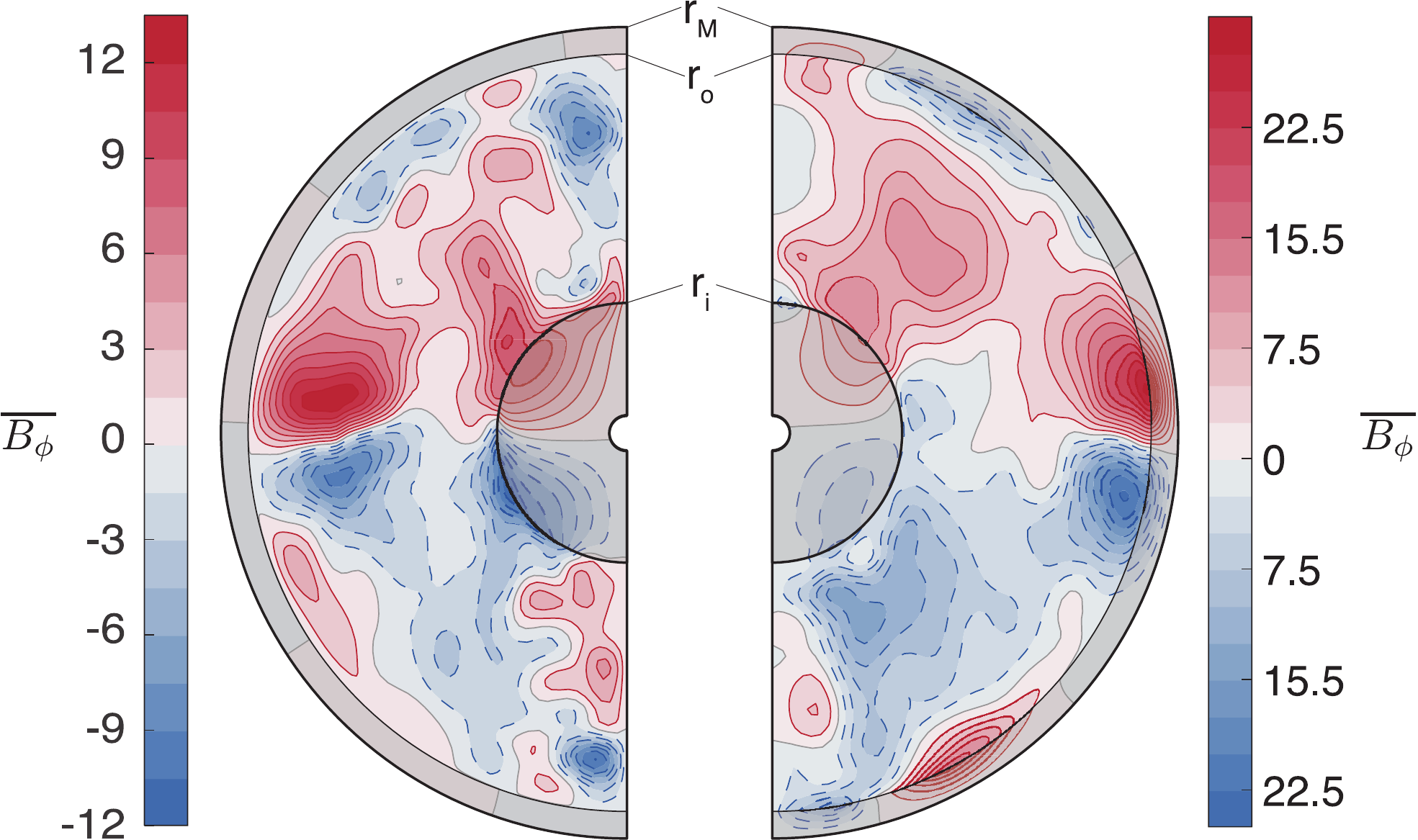}
\caption{Contours of the axisymmetric azimuthal ($\phi$) component of the magnetic field in the core for an insulating mantle model  ($\sigma_{MC}=1/400$, left) and for a conducting mantle model ($\sigma_{MC}=1$, right) at a single moment in time. The shaded regions indicate the inner core (center) and the conducting mantle layer (outside). Note the difference in scales between the two plots. In these models $r_{io}=0.35$.}
\label{fig:toroidal}
\end{figure}

In all cases we find that the addition of an electrically conducting mantle significantly enhances magnetic field generation in our models. The most striking example of this is in the axisymmetric azimuthal ($\phi$) component of the field near the CMB (Figure \ref{fig:toroidal}). Here magnetic fields anchored in both the solid, stationary mantle, and the fluid outer core are sheared by the strong zonal flows at the top of the core. This provides a source of magnetic field stretching, which strengthens the magnetic field in the azimuthal direction.

\begin{figure}
\epsscale{1}
\plotone{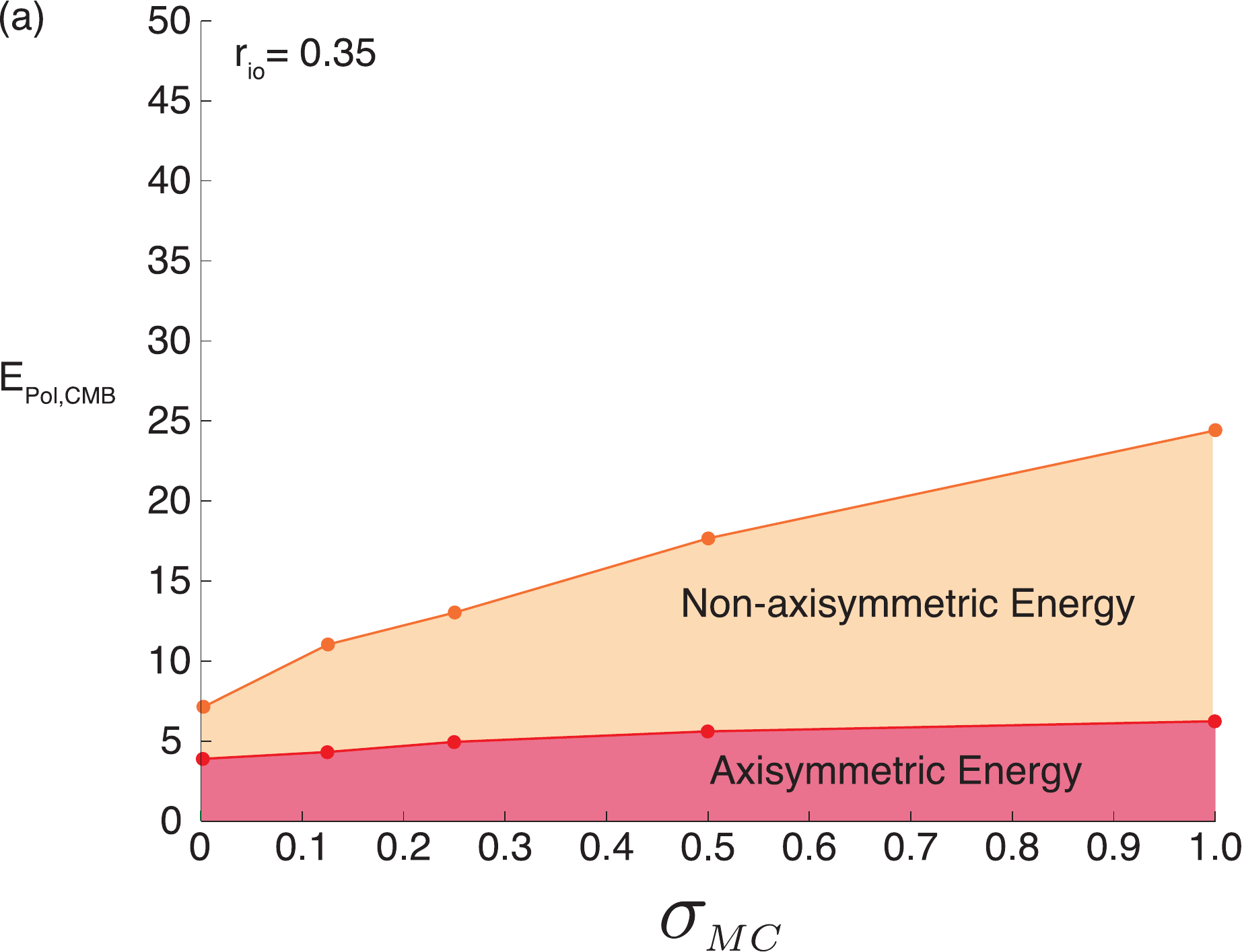}
\epsscale{1}
\plotone{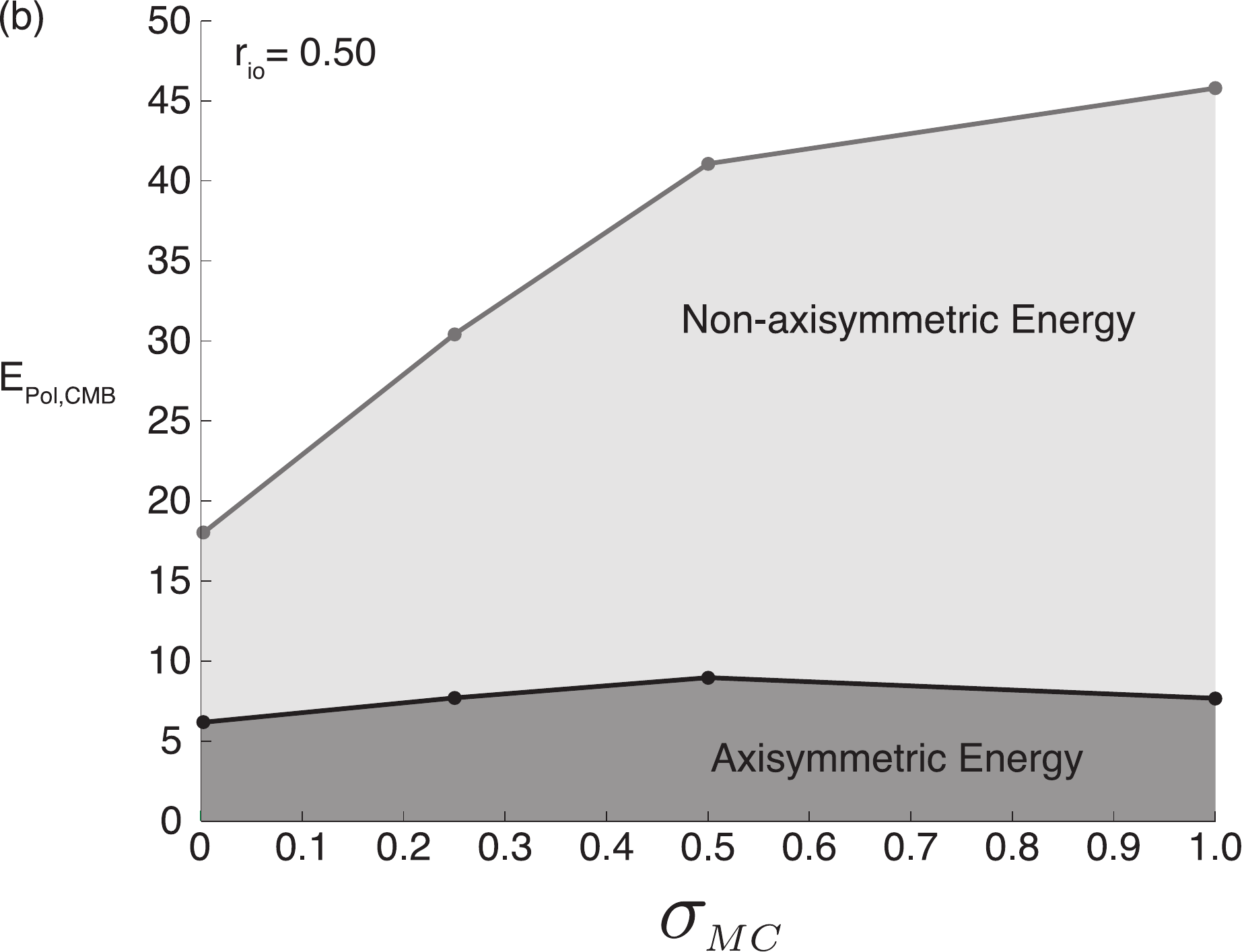}
\epsscale{1}
\plotone{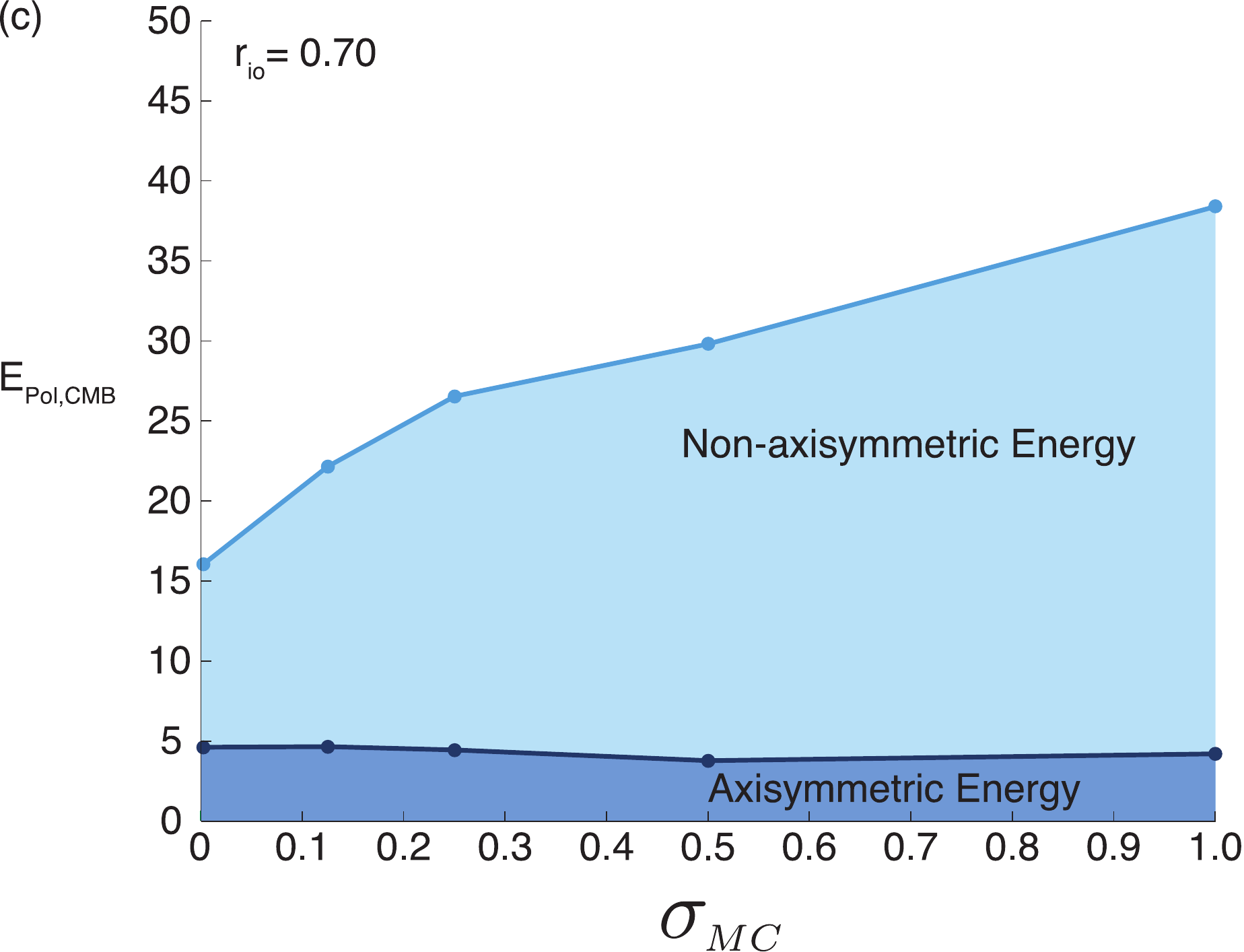}
\caption{Poloidal energies at the core-mantle boundary for models with $r_{io}=0.35$ (a), $r_{io}=0.50$ (b), and $r_{io}=0.70$ (c) separated into non-axisymmetric components (upper) and axisymmetric components (lower). All points are a time average over at least one magnetic diffusion time.} 
\label{fig:cmbenergies}
\end{figure}

We find that the strength of the poloidal component of the field at the CMB is increased as well (Figure \ref{fig:cmbenergies}). The poloidal component of the field is of special interest as it is the component which is observable from outside the conducting region (the toroidal field requires poloidal currents, which are only present in the conducting region). In our models, the energy in the non-axisymmetric component of the poloidal field increases as the conductivity of the mantle layer increases. In all cases the energy in the axisymmetric component of the poloidal field remains approximately constant.

This preferential increase in the non-axisymmetric component of the field can be explained by noting that the toroidal field at the CMB is predominantly axisymmetric, due to the large zonal axisymmetric flows shearing magnetic field there. Because a dynamo cannot create axisymmetric poloidal energy by any flows acting on axisymmetric toroidal magnetic fields \citep{bullard1954}, we should expect that the axisymmetric poloidal field should not increase due to an increase in the axisymmetric toroidal field caused by a conducting mantle layer. It is possible to create non-axisymmetric poloidal fields via non-axisymmetric velocities acting on axisymmetric toroidal magnetic fields, so an increase in the strength of the axisymmetric toroidal magnetic field should imply an increase in the strength of the non-axisymmetric poloidal magnetic field. 

\begin{figure}
\plotone{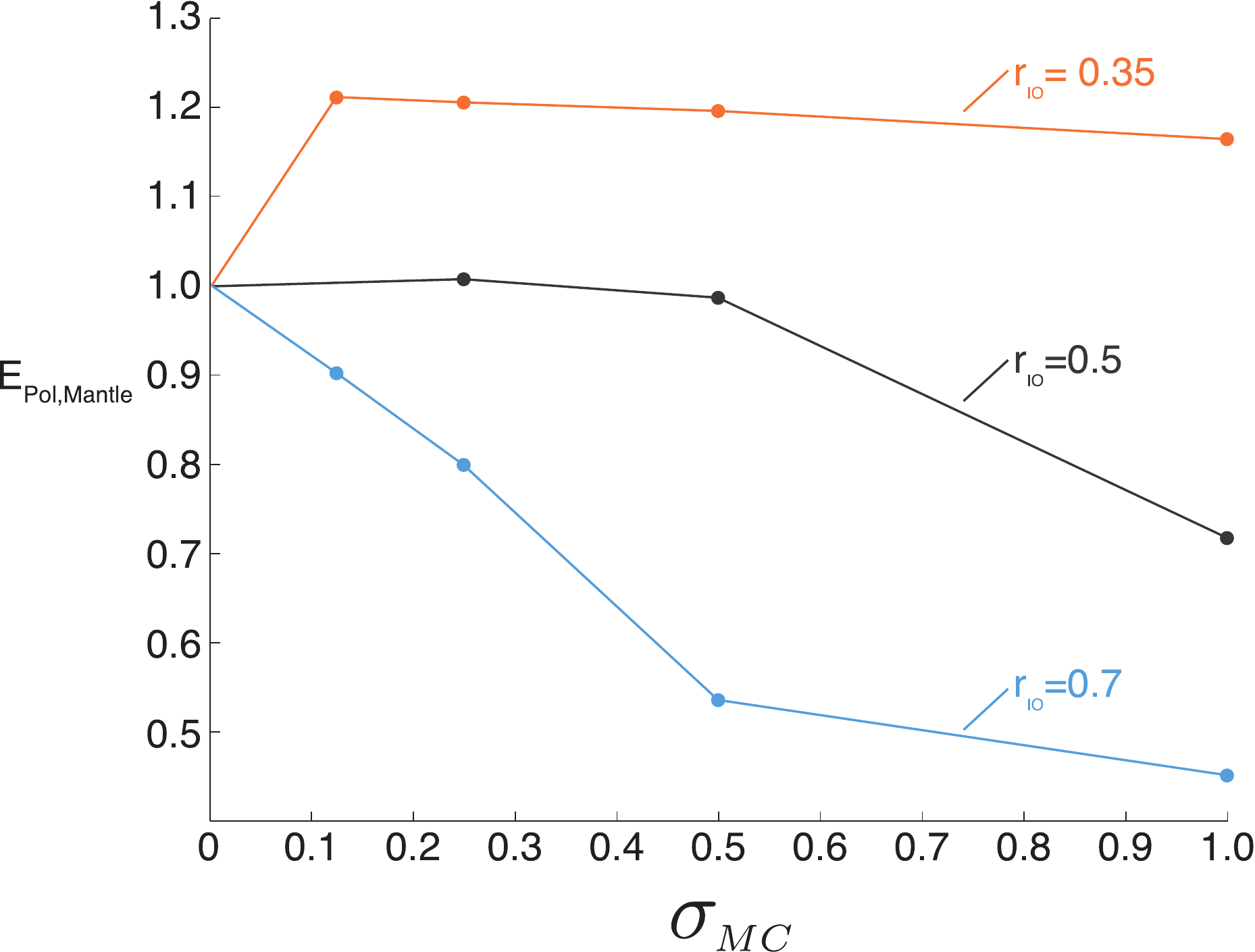}
\caption{Total poloidal energy at the top of the conducting layer ($r_{M}$) as a function of mantle conductivity. All points are a time average over at least one magnetic diffusion time and have been normalized to the insulating mantle case.}
\label{fig:ddppoloidaltotal}
\end{figure}

The axisymmetry of a dynamo is an important factor in the potential observability of extrasolar terrestrial planetary magnetic fields. For the magnetic field to reach the exterior of the planet where it can be observed, it must first be screened through a conducting lower mantle. As  discussed earlier, non-axisymmetric fields are screened more efficiently than axisymmetric fields. This becomes apparent if we plot the poloidal magnetic energies above the conducting layer (Figure \ref{fig:ddppoloidaltotal}). We see that depending on the shell thickness, the observable field is either only marginally stronger or much weaker than the field in a model without a conducting mantle layer.

As $r_{io}$ increases there is a marked decrease in poloidal field strength at the top of the conducting layer relative to the insulating case (Figure \ref{fig:ddppoloidaltotal}). When considered with Figure \ref{fig:cmbenergies}, the reason for this becomes clear. As the liquid outer core becomes thinner the dynamo becomes less axisymmetric at the top of the dynamo region and the characteristic timescale of variation becomes shorter \citep{aubert2009}. Both of these effects cause the field to be weakened by the screening effect of the conducting layer. This means that thinner shells are more susceptible to the screening effect discussed earlier. The increase in non-axisymmetry with increasing $r_{io}$ has been observed previously in studies which modelled the evolution of the Earth's dynamo through time \citep{aubert2009, roberts2001}.

\section{Conclusions}

We have used a numerical planetary dynamo model to investigate the effect of the metallization of silicate mantles on the observable magnetic fields of super-Earths. We have carried out models for three different inner core sizes to simulate these planets at different stages in their thermal evolution. In all cases we find that the strength of the internal magnetic field increases substantially, owing to the magnetic shear provided at the CMB by the conducting mantle. We also find that the addition of a conducting mantle makes the field significantly less axisymmetric at the top of the dynamo region. After being screened through the conducting mantle layer we find that the observable field shows either a modest increase in field strength (at $r_{io}=0.35$) or a significant decrease in field strength (at $r_{io}=0.7$).

As we have used a thin conducting layer in these models (compared to the range that is possible for terrestrial exoplanets), we expect that in larger planets, the screening effect would be even stronger than we observe here. This means that any planets with a metallized mantle should have surface fields which have been significantly weakened by a combination of the non-axisymmetrization of the dynamo, and the screening effect of the mantle. We therefore expect that the metallization of silicates should make the detection of dynamo-generated magnetic fields from super-Earth's more difficult than previously anticipated \citep{driscoll2011}.

\section{Acknowledgments}
The authors thank two anonymous reviewers for useful comments that improved this manuscript. S. Stanley acknowledges funding for this project by the National Research Council of Canada (NSERC) and the Alfred P. Sloan Foundation. Simulations were performed on the gpc and tcs supercomputers at the SciNet HPC Consortium. L. Elkins-Tanton acknowledges funding for this project from an NSF Astronomy CAREER award, NSF 6917282.

\end{document}